\documentclass[smallextended]{svjour3}
\smartqed
\usepackage{graphicx}
\begin{document}
\title{Correlated Gaussians and low-discrepancy sequences}
\author{D.V. Fedorov}
\institute{D.V. Fedorov \at
   Institute of Physics and Astronomy \\
   Aarhus University \\
   Ny Munkegade 120 \\
   8000 Aarhus C
   \email{fedorov@phys.au.dk}
}

\date{}
\maketitle
\begin{abstract}
Within the Correlated Gaussian Method
the parameters of the Gaussian basis functions
are often chosen stochastically using pseudo-random
sequences.
We show that alternative low-discrepancy sequences, also
known as quasi-random sequences, provide bases of better quality.

\end{abstract}

\section{Introduction}
The Correlated Gaussian Method is a popular variational method for
solving few-body problems in quantum mechanics~\cite{suzuki-varga,mitroy}.
The wave-function of a quantum few-body system is represented as a linear
combination of correlated Gaussians.  This reduces the corresponding
Schrodinger equation to a generalised matrix eigenvalue
problem in the (non-orthogonal) basis of correlated Gaussians.  One of
the advantages of the Correlated Gaussian Method is that the involved
matrix elements are often analytic~\cite{analytic} which significantly
simplifies the calculations.

In realistic calculations the dimension of the parameter space of the
Gaussian basis is typically too large to be sampled deterministically.
Therefore the basis is often constructed stochastically: the parameters of
the Gaussians in the basis are chosen randomly using a pseudo-random
number generator~\cite{suzuki-varga,mitroy}.

However, the pseudo-random sequences have the property of high discrepancy
or, in other words, low uniformity~\cite{morokoff}. Pseudo-random
sequences tend to produce relatively large sub-areas of the parameter
space that are not sampled at all, and other sub-areas that are sampled
excessively.  This can lead to bases of lower quality.

On the contrary, the low-discrepancy sequences (also called quasi-random
sequences) are designed with the specific purpose to reduce discrepancy
and to sample uniformly all sub-areas.  Quasi-random sequences might
thus potentially produce bases of higher quality.

We compare the quality of Gaussian bases constructed stochastically using
pseudo- and quasi-random (Van der Corput) sequences by calculating the
ground state energies of two Coulombic three-body systems.  We show that
quasi-random sequences consistently outperform pseudo-random sequences by
providing a lower variational estimate of the ground-state energy.

\section{Pseudo-random and quasi-random sequences}

Pseudo-random sequences are computer generated sequences of numbers
which satisfy the statistical criteria for a truly random sequence,
that is, the absence of correlations of any type.  The consequence is
that pseudo-random sequences have high discrepancy or, equivalently,
low uniformity.  Pseudo-random sequences produce relatively large
non-sampled areas of the size on the order of $1/\sqrt{N}$, where $N$ is
the length of the sequence.  On the contrary, quasi-random sequences are
highly correlated sequences built with the specific purpose of reducing
the discrepancy down to approximately $1/N$~\cite{morokoff}.

Van der Corput sequence is one simple low-discrepancy sequence
constructed by reversing the base-$b$ representation of the sequence of
natural numbers~\cite{corput}.	If the $n$-th natural number in base-$b$
representation is given as
	\begin{equation}
n=\sum_{k=0}^{L-1}d_k(n)b^k \;,
	\end{equation}
where $d_k(n)$ are the digits, then the $n$-th number $q_n$ of
the van der Corput sequence is given as
	\begin{equation}
q_n=\sum_{k=0}^{L-1}d_k(n)b^{-k-1} \;.
	\end{equation}
The numbers in the van der Corput sequence are uniformly distributed
over the unit interval with the discrepancy of $O(\frac{\log{N}}{N})$.

\begin{figure}
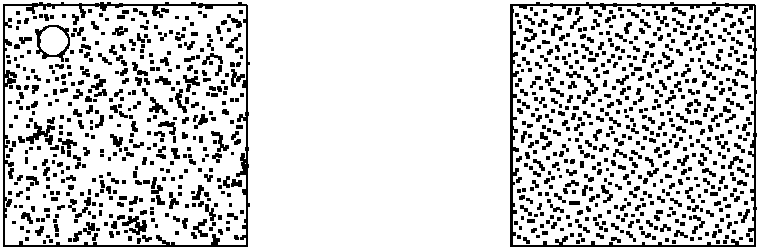

\input{pseudo.tex}
\input{halton.tex}
\caption{The first 1000 points of a two-dimensional unit-square
pseudo-random sequence (left) and quasi-random base-2/3 Van der Corput
sequence (right).  The pseudo-random sequence exhibits high
discrepancy: there exist relatively large areas,
exemplified by a circle at the upper left corner, with no points.
The quasi-random sequence distributes points with lower
discrepancy, that is, more uniformly.}
	\label{random} \end{figure}

Figure~(\ref{random}) illustrates the reduced discrepancy of the Van der
Corput sequence compared to a pseudo-random sequence.

\section{Examples}

We use two non-relativistic Coulombic three-body systems to compare
the relative performance of pseudo- and quasi-random sequences: the
positronium anion, $\mathrm{Ps}^-$, and the $td\mu^-$
anion.  We calculate the ground-state energies of these systems using
correlated Gaussian bases constructed stochastically with pseudo-random
and quasi-random sequences.

The Hamiltonian of a system of three bodies with massed $m_i$, charges
$q_i$, and cartesian coordinates $\vec r_i$, $i=1\dots3$, is given as
	\begin{equation}
H= -\sum_{i=1}^{3}\frac{1}{2m_i}\frac{\partial^2}{\partial\vec r_i^2}
+\sum_{i<j=1}^{3}\frac{q_iq_j}{|\vec r_i-\vec r_j|} \;,
	\end{equation}
where atomic units are used, that is, the unit of charge is the charge
of the positron, $e$, the unit of mass is the electron mass, $m_e$, the
unit of length is the Bohr radius, $a=\hbar^2/(m_ee^2)$, and the unit of
energy energy is Hartree, $h=m_ee^4/\hbar^2$.

For the positronium anion the masses of the three constituents are
$m_i=\{1,1,1\}$ and the charges $q_i=\{+1,-1,-1\}$.
For the $td\mu$ molecule we use $m_i=\{5496.918,3670.481,206.7686\}$
and $q_i=\{+1,+1,-1\}$~\cite{suzuki-varga}.

The correlated Gaussians are parameterised in the form
\begin{equation}
\exp\left(
-\sum_{i<j=1}^{3}
\left(\frac{\vec r_i-\vec r_j}{b_{ij}}\right)^2
\right) \;,
\end{equation}
where $b_{ij}$ are the range parameters of the Gaussians. The range
parameters are chosen stochastically following the
usual uniform sampling strategy~\cite{suzuki-varga},
	\begin{equation}\label{eq-u}
b_{ij}=u b_0 \;,
	\end{equation}
where the
number $u\in]0,1[$ is taken from a pseudo- or a quasi-random sequence,
and where $b_0$ is the appropriate scale factor on the order of the
size of the system.

For pseudo-random calculations we used the sequence produced by the
{\tt rand()} function from the Linux C~library seeded to 13.  For the
quasi-random calculations we used three Van der Corput sequences
with bases 2,~3, and~ 5.

The scale factor $b_0$ must be fine-tuned for a
particular calculation. An example of fine-tuning is shown on
figure~(\ref{fig-scale}) where the ground state energies of the two
systems are plotted as function of the scale factor for the two
sequences.  The figure shows that i) the minima at the
optimal values are relatively broad and need not to be calculated exactly,
and ii) the quasi-random sequence outperforms the pseudo-random sequence
for both systems.

We now take the optimum values of the scale factor from
figure~(\ref{fig-scale}) and calculate the ground state energies of the two
systems for bases with different sizes. Although the optimum scale factors
depend slightly on the size of the basis, the dependence is not strong and
can be neglected for the variations of the sizes that we use.

\begin{figure}
\setlength{\unitlength}{0.240900pt}
\ifx\plotpoint\undefined\newsavebox{\plotpoint}\fi
\sbox{\plotpoint}{\rule[-0.200pt]{0.400pt}{0.400pt}}%
\begin{picture}(705,630)(0,0)
\sbox{\plotpoint}{\rule[-0.200pt]{0.400pt}{0.400pt}}%
\put(211.0,131.0){\rule[-0.200pt]{4.818pt}{0.400pt}}
\put(191,131){\makebox(0,0)[r]{$0.0001$}}
\put(624.0,131.0){\rule[-0.200pt]{4.818pt}{0.400pt}}
\put(211.0,187.0){\rule[-0.200pt]{2.409pt}{0.400pt}}
\put(634.0,187.0){\rule[-0.200pt]{2.409pt}{0.400pt}}
\put(211.0,220.0){\rule[-0.200pt]{2.409pt}{0.400pt}}
\put(634.0,220.0){\rule[-0.200pt]{2.409pt}{0.400pt}}
\put(211.0,244.0){\rule[-0.200pt]{2.409pt}{0.400pt}}
\put(634.0,244.0){\rule[-0.200pt]{2.409pt}{0.400pt}}
\put(211.0,262.0){\rule[-0.200pt]{2.409pt}{0.400pt}}
\put(634.0,262.0){\rule[-0.200pt]{2.409pt}{0.400pt}}
\put(211.0,277.0){\rule[-0.200pt]{2.409pt}{0.400pt}}
\put(634.0,277.0){\rule[-0.200pt]{2.409pt}{0.400pt}}
\put(211.0,289.0){\rule[-0.200pt]{2.409pt}{0.400pt}}
\put(634.0,289.0){\rule[-0.200pt]{2.409pt}{0.400pt}}
\put(211.0,300.0){\rule[-0.200pt]{2.409pt}{0.400pt}}
\put(634.0,300.0){\rule[-0.200pt]{2.409pt}{0.400pt}}
\put(211.0,310.0){\rule[-0.200pt]{2.409pt}{0.400pt}}
\put(634.0,310.0){\rule[-0.200pt]{2.409pt}{0.400pt}}
\put(211.0,319.0){\rule[-0.200pt]{4.818pt}{0.400pt}}
\put(191,319){\makebox(0,0)[r]{$0.001$}}
\put(624.0,319.0){\rule[-0.200pt]{4.818pt}{0.400pt}}
\put(211.0,375.0){\rule[-0.200pt]{2.409pt}{0.400pt}}
\put(634.0,375.0){\rule[-0.200pt]{2.409pt}{0.400pt}}
\put(211.0,408.0){\rule[-0.200pt]{2.409pt}{0.400pt}}
\put(634.0,408.0){\rule[-0.200pt]{2.409pt}{0.400pt}}
\put(211.0,431.0){\rule[-0.200pt]{2.409pt}{0.400pt}}
\put(634.0,431.0){\rule[-0.200pt]{2.409pt}{0.400pt}}
\put(211.0,450.0){\rule[-0.200pt]{2.409pt}{0.400pt}}
\put(634.0,450.0){\rule[-0.200pt]{2.409pt}{0.400pt}}
\put(211.0,464.0){\rule[-0.200pt]{2.409pt}{0.400pt}}
\put(634.0,464.0){\rule[-0.200pt]{2.409pt}{0.400pt}}
\put(211.0,477.0){\rule[-0.200pt]{2.409pt}{0.400pt}}
\put(634.0,477.0){\rule[-0.200pt]{2.409pt}{0.400pt}}
\put(211.0,488.0){\rule[-0.200pt]{2.409pt}{0.400pt}}
\put(634.0,488.0){\rule[-0.200pt]{2.409pt}{0.400pt}}
\put(211.0,497.0){\rule[-0.200pt]{2.409pt}{0.400pt}}
\put(634.0,497.0){\rule[-0.200pt]{2.409pt}{0.400pt}}
\put(211.0,506.0){\rule[-0.200pt]{4.818pt}{0.400pt}}
\put(191,506){\makebox(0,0)[r]{$0.01$}}
\put(624.0,506.0){\rule[-0.200pt]{4.818pt}{0.400pt}}
\put(211.0,131.0){\rule[-0.200pt]{0.400pt}{4.818pt}}
\put(211,90){\makebox(0,0){$10$}}
\put(211.0,486.0){\rule[-0.200pt]{0.400pt}{4.818pt}}
\put(283.0,131.0){\rule[-0.200pt]{0.400pt}{4.818pt}}
\put(283,90){\makebox(0,0){$15$}}
\put(283.0,486.0){\rule[-0.200pt]{0.400pt}{4.818pt}}
\put(355.0,131.0){\rule[-0.200pt]{0.400pt}{4.818pt}}
\put(355,90){\makebox(0,0){$20$}}
\put(355.0,486.0){\rule[-0.200pt]{0.400pt}{4.818pt}}
\put(428.0,131.0){\rule[-0.200pt]{0.400pt}{4.818pt}}
\put(428,90){\makebox(0,0){$25$}}
\put(428.0,486.0){\rule[-0.200pt]{0.400pt}{4.818pt}}
\put(500.0,131.0){\rule[-0.200pt]{0.400pt}{4.818pt}}
\put(500,90){\makebox(0,0){$30$}}
\put(500.0,486.0){\rule[-0.200pt]{0.400pt}{4.818pt}}
\put(572.0,131.0){\rule[-0.200pt]{0.400pt}{4.818pt}}
\put(572,90){\makebox(0,0){$35$}}
\put(572.0,486.0){\rule[-0.200pt]{0.400pt}{4.818pt}}
\put(644.0,131.0){\rule[-0.200pt]{0.400pt}{4.818pt}}
\put(644,90){\makebox(0,0){$40$}}
\put(644.0,486.0){\rule[-0.200pt]{0.400pt}{4.818pt}}
\put(211.0,131.0){\rule[-0.200pt]{0.400pt}{90.337pt}}
\put(211.0,131.0){\rule[-0.200pt]{104.310pt}{0.400pt}}
\put(644.0,131.0){\rule[-0.200pt]{0.400pt}{90.337pt}}
\put(211.0,506.0){\rule[-0.200pt]{104.310pt}{0.400pt}}
\put(30,318){\rotatebox{-270}{\makebox(0,0){$\frac{E-E_x}{|E_x|}$}}
}\put(427,29){\makebox(0,0){scale factor $ b_0 $}}
\put(427,568){\makebox(0,0){$ E $[Ps$ ^- $], 300 gaussians}}
\put(484,465){\makebox(0,0)[r]{pseudo}}
\put(504.0,465.0){\rule[-0.200pt]{24.090pt}{0.400pt}}
\put(283,340){\usebox{\plotpoint}}
\multiput(283.58,337.58)(0.497,-0.603){55}{\rule{0.120pt}{0.583pt}}
\multiput(282.17,338.79)(29.000,-33.790){2}{\rule{0.400pt}{0.291pt}}
\multiput(312.00,303.92)(0.660,-0.496){41}{\rule{0.627pt}{0.120pt}}
\multiput(312.00,304.17)(27.698,-22.000){2}{\rule{0.314pt}{0.400pt}}
\multiput(341.00,281.93)(2.171,-0.485){11}{\rule{1.757pt}{0.117pt}}
\multiput(341.00,282.17)(25.353,-7.000){2}{\rule{0.879pt}{0.400pt}}
\multiput(370.00,276.59)(2.570,0.482){9}{\rule{2.033pt}{0.116pt}}
\multiput(370.00,275.17)(24.780,6.000){2}{\rule{1.017pt}{0.400pt}}
\multiput(399.00,282.58)(1.229,0.492){21}{\rule{1.067pt}{0.119pt}}
\multiput(399.00,281.17)(26.786,12.000){2}{\rule{0.533pt}{0.400pt}}
\multiput(428.00,294.58)(0.942,0.494){27}{\rule{0.847pt}{0.119pt}}
\multiput(428.00,293.17)(26.243,15.000){2}{\rule{0.423pt}{0.400pt}}
\multiput(456.00,309.58)(0.976,0.494){27}{\rule{0.873pt}{0.119pt}}
\multiput(456.00,308.17)(27.187,15.000){2}{\rule{0.437pt}{0.400pt}}
\sbox{\plotpoint}{\rule[-0.500pt]{1.000pt}{1.000pt}}%
\sbox{\plotpoint}{\rule[-0.200pt]{0.400pt}{0.400pt}}%
\put(484,424){\makebox(0,0)[r]{quasi}}
\sbox{\plotpoint}{\rule[-0.500pt]{1.000pt}{1.000pt}}%
\multiput(504,424)(20.756,0.000){5}{\usebox{\plotpoint}}
\put(604,424){\usebox{\plotpoint}}
\put(355,242){\usebox{\plotpoint}}
\multiput(355,242)(14.179,-15.157){3}{\usebox{\plotpoint}}
\put(399.53,198.69){\usebox{\plotpoint}}
\multiput(413,188)(18.940,-8.490){2}{\usebox{\plotpoint}}
\put(455.26,174.09){\usebox{\plotpoint}}
\multiput(471,173)(20.176,4.870){2}{\usebox{\plotpoint}}
\put(515.20,186.81){\usebox{\plotpoint}}
\multiput(529,193)(17.742,10.772){2}{\usebox{\plotpoint}}
\put(557,210){\usebox{\plotpoint}}
\sbox{\plotpoint}{\rule[-0.200pt]{0.400pt}{0.400pt}}%
\put(211.0,131.0){\rule[-0.200pt]{0.400pt}{90.337pt}}
\put(211.0,131.0){\rule[-0.200pt]{104.310pt}{0.400pt}}
\put(644.0,131.0){\rule[-0.200pt]{0.400pt}{90.337pt}}
\put(211.0,506.0){\rule[-0.200pt]{104.310pt}{0.400pt}}
\end{picture}
\setlength{\unitlength}{0.240900pt}
\ifx\plotpoint\undefined\newsavebox{\plotpoint}\fi
\sbox{\plotpoint}{\rule[-0.200pt]{0.400pt}{0.400pt}}%
\begin{picture}(705,630)(0,0)
\sbox{\plotpoint}{\rule[-0.200pt]{0.400pt}{0.400pt}}%
\put(211.0,131.0){\rule[-0.200pt]{4.818pt}{0.400pt}}
\put(191,131){\makebox(0,0)[r]{$0.0001$}}
\put(624.0,131.0){\rule[-0.200pt]{4.818pt}{0.400pt}}
\put(211.0,187.0){\rule[-0.200pt]{2.409pt}{0.400pt}}
\put(634.0,187.0){\rule[-0.200pt]{2.409pt}{0.400pt}}
\put(211.0,220.0){\rule[-0.200pt]{2.409pt}{0.400pt}}
\put(634.0,220.0){\rule[-0.200pt]{2.409pt}{0.400pt}}
\put(211.0,244.0){\rule[-0.200pt]{2.409pt}{0.400pt}}
\put(634.0,244.0){\rule[-0.200pt]{2.409pt}{0.400pt}}
\put(211.0,262.0){\rule[-0.200pt]{2.409pt}{0.400pt}}
\put(634.0,262.0){\rule[-0.200pt]{2.409pt}{0.400pt}}
\put(211.0,277.0){\rule[-0.200pt]{2.409pt}{0.400pt}}
\put(634.0,277.0){\rule[-0.200pt]{2.409pt}{0.400pt}}
\put(211.0,289.0){\rule[-0.200pt]{2.409pt}{0.400pt}}
\put(634.0,289.0){\rule[-0.200pt]{2.409pt}{0.400pt}}
\put(211.0,300.0){\rule[-0.200pt]{2.409pt}{0.400pt}}
\put(634.0,300.0){\rule[-0.200pt]{2.409pt}{0.400pt}}
\put(211.0,310.0){\rule[-0.200pt]{2.409pt}{0.400pt}}
\put(634.0,310.0){\rule[-0.200pt]{2.409pt}{0.400pt}}
\put(211.0,319.0){\rule[-0.200pt]{4.818pt}{0.400pt}}
\put(191,319){\makebox(0,0)[r]{$0.001$}}
\put(624.0,319.0){\rule[-0.200pt]{4.818pt}{0.400pt}}
\put(211.0,375.0){\rule[-0.200pt]{2.409pt}{0.400pt}}
\put(634.0,375.0){\rule[-0.200pt]{2.409pt}{0.400pt}}
\put(211.0,408.0){\rule[-0.200pt]{2.409pt}{0.400pt}}
\put(634.0,408.0){\rule[-0.200pt]{2.409pt}{0.400pt}}
\put(211.0,431.0){\rule[-0.200pt]{2.409pt}{0.400pt}}
\put(634.0,431.0){\rule[-0.200pt]{2.409pt}{0.400pt}}
\put(211.0,450.0){\rule[-0.200pt]{2.409pt}{0.400pt}}
\put(634.0,450.0){\rule[-0.200pt]{2.409pt}{0.400pt}}
\put(211.0,464.0){\rule[-0.200pt]{2.409pt}{0.400pt}}
\put(634.0,464.0){\rule[-0.200pt]{2.409pt}{0.400pt}}
\put(211.0,477.0){\rule[-0.200pt]{2.409pt}{0.400pt}}
\put(634.0,477.0){\rule[-0.200pt]{2.409pt}{0.400pt}}
\put(211.0,488.0){\rule[-0.200pt]{2.409pt}{0.400pt}}
\put(634.0,488.0){\rule[-0.200pt]{2.409pt}{0.400pt}}
\put(211.0,497.0){\rule[-0.200pt]{2.409pt}{0.400pt}}
\put(634.0,497.0){\rule[-0.200pt]{2.409pt}{0.400pt}}
\put(211.0,506.0){\rule[-0.200pt]{4.818pt}{0.400pt}}
\put(191,506){\makebox(0,0)[r]{$0.01$}}
\put(624.0,506.0){\rule[-0.200pt]{4.818pt}{0.400pt}}
\put(211.0,131.0){\rule[-0.200pt]{0.400pt}{4.818pt}}
\put(211,90){\makebox(0,0){$0.02$}}
\put(211.0,486.0){\rule[-0.200pt]{0.400pt}{4.818pt}}
\put(319.0,131.0){\rule[-0.200pt]{0.400pt}{4.818pt}}
\put(319,90){\makebox(0,0){$0.025$}}
\put(319.0,486.0){\rule[-0.200pt]{0.400pt}{4.818pt}}
\put(428.0,131.0){\rule[-0.200pt]{0.400pt}{4.818pt}}
\put(428,90){\makebox(0,0){$0.03$}}
\put(428.0,486.0){\rule[-0.200pt]{0.400pt}{4.818pt}}
\put(536.0,131.0){\rule[-0.200pt]{0.400pt}{4.818pt}}
\put(536,90){\makebox(0,0){$0.035$}}
\put(536.0,486.0){\rule[-0.200pt]{0.400pt}{4.818pt}}
\put(644.0,131.0){\rule[-0.200pt]{0.400pt}{4.818pt}}
\put(644,90){\makebox(0,0){$0.04$}}
\put(644.0,486.0){\rule[-0.200pt]{0.400pt}{4.818pt}}
\put(211.0,131.0){\rule[-0.200pt]{0.400pt}{90.337pt}}
\put(211.0,131.0){\rule[-0.200pt]{104.310pt}{0.400pt}}
\put(644.0,131.0){\rule[-0.200pt]{0.400pt}{90.337pt}}
\put(211.0,506.0){\rule[-0.200pt]{104.310pt}{0.400pt}}
\put(30,318){\rotatebox{-270}{\makebox(0,0){$\frac{E-E_x}{|E_x|}$}}
}\put(427,29){\makebox(0,0){scale factor $ b_0 $}}
\put(427,568){\makebox(0,0){$ E[td\mu] $, 180 gaussians}}
\put(484,465){\makebox(0,0)[r]{pseudo}}
\put(504.0,465.0){\rule[-0.200pt]{24.090pt}{0.400pt}}
\put(276,252){\usebox{\plotpoint}}
\multiput(276.00,250.93)(0.762,-0.482){9}{\rule{0.700pt}{0.116pt}}
\multiput(276.00,251.17)(7.547,-6.000){2}{\rule{0.350pt}{0.400pt}}
\multiput(285.00,244.93)(0.671,-0.482){9}{\rule{0.633pt}{0.116pt}}
\multiput(285.00,245.17)(6.685,-6.000){2}{\rule{0.317pt}{0.400pt}}
\multiput(293.00,238.94)(1.212,-0.468){5}{\rule{1.000pt}{0.113pt}}
\multiput(293.00,239.17)(6.924,-4.000){2}{\rule{0.500pt}{0.400pt}}
\multiput(302.00,234.94)(1.212,-0.468){5}{\rule{1.000pt}{0.113pt}}
\multiput(302.00,235.17)(6.924,-4.000){2}{\rule{0.500pt}{0.400pt}}
\put(311,230.17){\rule{1.700pt}{0.400pt}}
\multiput(311.00,231.17)(4.472,-2.000){2}{\rule{0.850pt}{0.400pt}}
\put(319,228.17){\rule{1.900pt}{0.400pt}}
\multiput(319.00,229.17)(5.056,-2.000){2}{\rule{0.950pt}{0.400pt}}
\multiput(337.00,226.95)(1.579,-0.447){3}{\rule{1.167pt}{0.108pt}}
\multiput(337.00,227.17)(5.579,-3.000){2}{\rule{0.583pt}{0.400pt}}
\put(328.0,228.0){\rule[-0.200pt]{2.168pt}{0.400pt}}
\multiput(363.00,225.61)(1.579,0.447){3}{\rule{1.167pt}{0.108pt}}
\multiput(363.00,224.17)(5.579,3.000){2}{\rule{0.583pt}{0.400pt}}
\put(345.0,225.0){\rule[-0.200pt]{4.336pt}{0.400pt}}
\put(380,228.17){\rule{1.900pt}{0.400pt}}
\multiput(380.00,227.17)(5.056,2.000){2}{\rule{0.950pt}{0.400pt}}
\put(389,230.17){\rule{1.700pt}{0.400pt}}
\multiput(389.00,229.17)(4.472,2.000){2}{\rule{0.850pt}{0.400pt}}
\put(397,232.17){\rule{1.900pt}{0.400pt}}
\multiput(397.00,231.17)(5.056,2.000){2}{\rule{0.950pt}{0.400pt}}
\put(406,234.17){\rule{1.900pt}{0.400pt}}
\multiput(406.00,233.17)(5.056,2.000){2}{\rule{0.950pt}{0.400pt}}
\put(415,236.17){\rule{1.700pt}{0.400pt}}
\multiput(415.00,235.17)(4.472,2.000){2}{\rule{0.850pt}{0.400pt}}
\put(423,238.17){\rule{1.900pt}{0.400pt}}
\multiput(423.00,237.17)(5.056,2.000){2}{\rule{0.950pt}{0.400pt}}
\put(432,240.17){\rule{1.700pt}{0.400pt}}
\multiput(432.00,239.17)(4.472,2.000){2}{\rule{0.850pt}{0.400pt}}
\multiput(440.00,242.60)(1.212,0.468){5}{\rule{1.000pt}{0.113pt}}
\multiput(440.00,241.17)(6.924,4.000){2}{\rule{0.500pt}{0.400pt}}
\put(449,245.67){\rule{2.168pt}{0.400pt}}
\multiput(449.00,245.17)(4.500,1.000){2}{\rule{1.084pt}{0.400pt}}
\multiput(458.00,247.60)(1.066,0.468){5}{\rule{0.900pt}{0.113pt}}
\multiput(458.00,246.17)(6.132,4.000){2}{\rule{0.450pt}{0.400pt}}
\put(371.0,228.0){\rule[-0.200pt]{2.168pt}{0.400pt}}
\sbox{\plotpoint}{\rule[-0.500pt]{1.000pt}{1.000pt}}%
\sbox{\plotpoint}{\rule[-0.200pt]{0.400pt}{0.400pt}}%
\put(484,424){\makebox(0,0)[r]{quasi}}
\sbox{\plotpoint}{\rule[-0.500pt]{1.000pt}{1.000pt}}%
\multiput(504,424)(20.756,0.000){5}{\usebox{\plotpoint}}
\put(604,424){\usebox{\plotpoint}}
\put(298,204){\usebox{\plotpoint}}
\put(298.00,204.00){\usebox{\plotpoint}}
\put(314.95,192.03){\usebox{\plotpoint}}
\put(333.74,183.23){\usebox{\plotpoint}}
\put(353.41,178.30){\usebox{\plotpoint}}
\put(373.62,176.00){\usebox{\plotpoint}}
\put(394.37,176.00){\usebox{\plotpoint}}
\put(414.19,180.00){\usebox{\plotpoint}}
\put(433.45,186.72){\usebox{\plotpoint}}
\put(452.96,191.98){\usebox{\plotpoint}}
\put(472.63,198.61){\usebox{\plotpoint}}
\put(488,207){\usebox{\plotpoint}}
\sbox{\plotpoint}{\rule[-0.200pt]{0.400pt}{0.400pt}}%
\put(211.0,131.0){\rule[-0.200pt]{0.400pt}{90.337pt}}
\put(211.0,131.0){\rule[-0.200pt]{104.310pt}{0.400pt}}
\put(644.0,131.0){\rule[-0.200pt]{0.400pt}{90.337pt}}
\put(211.0,506.0){\rule[-0.200pt]{104.310pt}{0.400pt}}
\end{picture}
\caption{The ground state energy $E$ of the Ps$^-$ (left)
and $td\mu$ (right) systems
as
function of the sampling scale factor $b_0$ from equation~(\ref{eq-u}).
The energies are calculated using
stochastically chosen correlated Gaussian
bases constructed with pseudo- or quasi-random sequences.
The exact energies are
$E_x=-0.26200507$ for Ps$^-$ and
$E_x=-111.36444$ for $td\mu$~\cite{suzuki-varga}.
		  }
	\label{fig-scale}
	\end{figure}
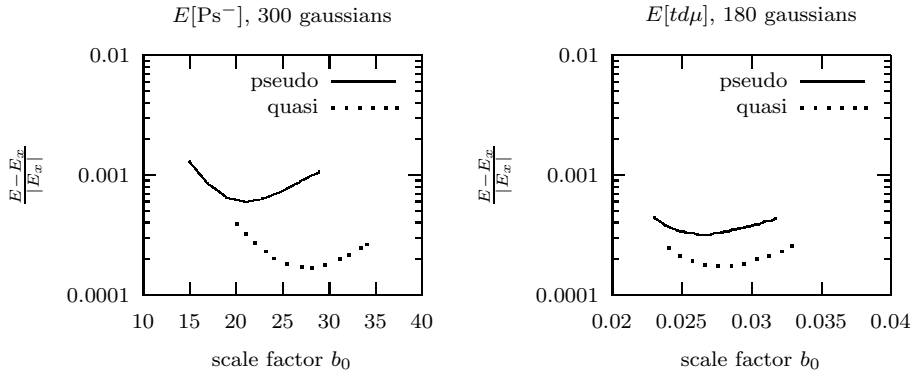

\begin{figure}
\setlength{\unitlength}{0.240900pt}
\ifx\plotpoint\undefined\newsavebox{\plotpoint}\fi
\sbox{\plotpoint}{\rule[-0.200pt]{0.400pt}{0.400pt}}%
\begin{picture}(705,630)(0,0)
\sbox{\plotpoint}{\rule[-0.200pt]{0.400pt}{0.400pt}}%
\put(211.0,131.0){\rule[-0.200pt]{4.818pt}{0.400pt}}
\put(191,131){\makebox(0,0)[r]{$0.0001$}}
\put(624.0,131.0){\rule[-0.200pt]{4.818pt}{0.400pt}}
\put(211.0,187.0){\rule[-0.200pt]{2.409pt}{0.400pt}}
\put(634.0,187.0){\rule[-0.200pt]{2.409pt}{0.400pt}}
\put(211.0,220.0){\rule[-0.200pt]{2.409pt}{0.400pt}}
\put(634.0,220.0){\rule[-0.200pt]{2.409pt}{0.400pt}}
\put(211.0,244.0){\rule[-0.200pt]{2.409pt}{0.400pt}}
\put(634.0,244.0){\rule[-0.200pt]{2.409pt}{0.400pt}}
\put(211.0,262.0){\rule[-0.200pt]{2.409pt}{0.400pt}}
\put(634.0,262.0){\rule[-0.200pt]{2.409pt}{0.400pt}}
\put(211.0,277.0){\rule[-0.200pt]{2.409pt}{0.400pt}}
\put(634.0,277.0){\rule[-0.200pt]{2.409pt}{0.400pt}}
\put(211.0,289.0){\rule[-0.200pt]{2.409pt}{0.400pt}}
\put(634.0,289.0){\rule[-0.200pt]{2.409pt}{0.400pt}}
\put(211.0,300.0){\rule[-0.200pt]{2.409pt}{0.400pt}}
\put(634.0,300.0){\rule[-0.200pt]{2.409pt}{0.400pt}}
\put(211.0,310.0){\rule[-0.200pt]{2.409pt}{0.400pt}}
\put(634.0,310.0){\rule[-0.200pt]{2.409pt}{0.400pt}}
\put(211.0,319.0){\rule[-0.200pt]{4.818pt}{0.400pt}}
\put(191,319){\makebox(0,0)[r]{$0.001$}}
\put(624.0,319.0){\rule[-0.200pt]{4.818pt}{0.400pt}}
\put(211.0,375.0){\rule[-0.200pt]{2.409pt}{0.400pt}}
\put(634.0,375.0){\rule[-0.200pt]{2.409pt}{0.400pt}}
\put(211.0,408.0){\rule[-0.200pt]{2.409pt}{0.400pt}}
\put(634.0,408.0){\rule[-0.200pt]{2.409pt}{0.400pt}}
\put(211.0,431.0){\rule[-0.200pt]{2.409pt}{0.400pt}}
\put(634.0,431.0){\rule[-0.200pt]{2.409pt}{0.400pt}}
\put(211.0,450.0){\rule[-0.200pt]{2.409pt}{0.400pt}}
\put(634.0,450.0){\rule[-0.200pt]{2.409pt}{0.400pt}}
\put(211.0,464.0){\rule[-0.200pt]{2.409pt}{0.400pt}}
\put(634.0,464.0){\rule[-0.200pt]{2.409pt}{0.400pt}}
\put(211.0,477.0){\rule[-0.200pt]{2.409pt}{0.400pt}}
\put(634.0,477.0){\rule[-0.200pt]{2.409pt}{0.400pt}}
\put(211.0,488.0){\rule[-0.200pt]{2.409pt}{0.400pt}}
\put(634.0,488.0){\rule[-0.200pt]{2.409pt}{0.400pt}}
\put(211.0,497.0){\rule[-0.200pt]{2.409pt}{0.400pt}}
\put(634.0,497.0){\rule[-0.200pt]{2.409pt}{0.400pt}}
\put(211.0,506.0){\rule[-0.200pt]{4.818pt}{0.400pt}}
\put(191,506){\makebox(0,0)[r]{$0.01$}}
\put(624.0,506.0){\rule[-0.200pt]{4.818pt}{0.400pt}}
\put(211.0,131.0){\rule[-0.200pt]{0.400pt}{4.818pt}}
\put(211,90){\makebox(0,0){$100$}}
\put(211.0,486.0){\rule[-0.200pt]{0.400pt}{4.818pt}}
\put(283.0,131.0){\rule[-0.200pt]{0.400pt}{4.818pt}}
\put(283,90){\makebox(0,0){$150$}}
\put(283.0,486.0){\rule[-0.200pt]{0.400pt}{4.818pt}}
\put(355.0,131.0){\rule[-0.200pt]{0.400pt}{4.818pt}}
\put(355,90){\makebox(0,0){$200$}}
\put(355.0,486.0){\rule[-0.200pt]{0.400pt}{4.818pt}}
\put(428.0,131.0){\rule[-0.200pt]{0.400pt}{4.818pt}}
\put(428,90){\makebox(0,0){$250$}}
\put(428.0,486.0){\rule[-0.200pt]{0.400pt}{4.818pt}}
\put(500.0,131.0){\rule[-0.200pt]{0.400pt}{4.818pt}}
\put(500,90){\makebox(0,0){$300$}}
\put(500.0,486.0){\rule[-0.200pt]{0.400pt}{4.818pt}}
\put(572.0,131.0){\rule[-0.200pt]{0.400pt}{4.818pt}}
\put(572,90){\makebox(0,0){$350$}}
\put(572.0,486.0){\rule[-0.200pt]{0.400pt}{4.818pt}}
\put(644.0,131.0){\rule[-0.200pt]{0.400pt}{4.818pt}}
\put(644,90){\makebox(0,0){$400$}}
\put(644.0,486.0){\rule[-0.200pt]{0.400pt}{4.818pt}}
\put(211.0,131.0){\rule[-0.200pt]{0.400pt}{90.337pt}}
\put(211.0,131.0){\rule[-0.200pt]{104.310pt}{0.400pt}}
\put(644.0,131.0){\rule[-0.200pt]{0.400pt}{90.337pt}}
\put(211.0,506.0){\rule[-0.200pt]{104.310pt}{0.400pt}}
\put(30,318){\rotatebox{-270}{\makebox(0,0){$\frac{E-E_x}{|E_x|}$}}
}\put(427,29){\makebox(0,0){basis size $ n $}}
\put(427,568){\makebox(0,0){$ E $[Ps$ ^- $]}}
\put(484,465){\makebox(0,0)[r]{pseudo}}
\put(504.0,465.0){\rule[-0.200pt]{24.090pt}{0.400pt}}
\put(211,406){\usebox{\plotpoint}}
\multiput(211.58,402.26)(0.494,-1.011){25}{\rule{0.119pt}{0.900pt}}
\multiput(210.17,404.13)(14.000,-26.132){2}{\rule{0.400pt}{0.450pt}}
\multiput(225.58,375.04)(0.494,-0.771){27}{\rule{0.119pt}{0.713pt}}
\multiput(224.17,376.52)(15.000,-21.519){2}{\rule{0.400pt}{0.357pt}}
\multiput(240.00,353.94)(1.943,-0.468){5}{\rule{1.500pt}{0.113pt}}
\multiput(240.00,354.17)(10.887,-4.000){2}{\rule{0.750pt}{0.400pt}}
\put(254,349.67){\rule{3.614pt}{0.400pt}}
\multiput(254.00,350.17)(7.500,-1.000){2}{\rule{1.807pt}{0.400pt}}
\multiput(269.00,348.92)(0.704,-0.491){17}{\rule{0.660pt}{0.118pt}}
\multiput(269.00,349.17)(12.630,-10.000){2}{\rule{0.330pt}{0.400pt}}
\multiput(298.00,338.93)(0.890,-0.488){13}{\rule{0.800pt}{0.117pt}}
\multiput(298.00,339.17)(12.340,-8.000){2}{\rule{0.400pt}{0.400pt}}
\multiput(312.00,330.93)(1.026,-0.485){11}{\rule{0.900pt}{0.117pt}}
\multiput(312.00,331.17)(12.132,-7.000){2}{\rule{0.450pt}{0.400pt}}
\multiput(326.00,323.94)(2.090,-0.468){5}{\rule{1.600pt}{0.113pt}}
\multiput(326.00,324.17)(11.679,-4.000){2}{\rule{0.800pt}{0.400pt}}
\put(341,319.67){\rule{3.373pt}{0.400pt}}
\multiput(341.00,320.17)(7.000,-1.000){2}{\rule{1.686pt}{0.400pt}}
\multiput(355.00,318.94)(2.090,-0.468){5}{\rule{1.600pt}{0.113pt}}
\multiput(355.00,319.17)(11.679,-4.000){2}{\rule{0.800pt}{0.400pt}}
\multiput(370.00,314.93)(1.026,-0.485){11}{\rule{0.900pt}{0.117pt}}
\multiput(370.00,315.17)(12.132,-7.000){2}{\rule{0.450pt}{0.400pt}}
\multiput(384.00,307.92)(0.625,-0.492){21}{\rule{0.600pt}{0.119pt}}
\multiput(384.00,308.17)(13.755,-12.000){2}{\rule{0.300pt}{0.400pt}}
\put(399,295.17){\rule{2.900pt}{0.400pt}}
\multiput(399.00,296.17)(7.981,-2.000){2}{\rule{1.450pt}{0.400pt}}
\multiput(413.00,293.95)(3.141,-0.447){3}{\rule{2.100pt}{0.108pt}}
\multiput(413.00,294.17)(10.641,-3.000){2}{\rule{1.050pt}{0.400pt}}
\put(428,290.67){\rule{3.373pt}{0.400pt}}
\multiput(428.00,291.17)(7.000,-1.000){2}{\rule{1.686pt}{0.400pt}}
\multiput(442.00,289.93)(1.489,-0.477){7}{\rule{1.220pt}{0.115pt}}
\multiput(442.00,290.17)(11.468,-5.000){2}{\rule{0.610pt}{0.400pt}}
\put(283.0,340.0){\rule[-0.200pt]{3.613pt}{0.400pt}}
\multiput(471.00,284.93)(0.890,-0.488){13}{\rule{0.800pt}{0.117pt}}
\multiput(471.00,285.17)(12.340,-8.000){2}{\rule{0.400pt}{0.400pt}}
\put(485,276.17){\rule{3.100pt}{0.400pt}}
\multiput(485.00,277.17)(8.566,-2.000){2}{\rule{1.550pt}{0.400pt}}
\put(500,274.67){\rule{3.373pt}{0.400pt}}
\multiput(500.00,275.17)(7.000,-1.000){2}{\rule{1.686pt}{0.400pt}}
\multiput(514.58,272.48)(0.494,-0.634){27}{\rule{0.119pt}{0.607pt}}
\multiput(513.17,273.74)(15.000,-17.741){2}{\rule{0.400pt}{0.303pt}}
\multiput(529.00,254.93)(0.890,-0.488){13}{\rule{0.800pt}{0.117pt}}
\multiput(529.00,255.17)(12.340,-8.000){2}{\rule{0.400pt}{0.400pt}}
\put(543,246.17){\rule{2.900pt}{0.400pt}}
\multiput(543.00,247.17)(7.981,-2.000){2}{\rule{1.450pt}{0.400pt}}
\multiput(557.00,244.92)(0.756,-0.491){17}{\rule{0.700pt}{0.118pt}}
\multiput(557.00,245.17)(13.547,-10.000){2}{\rule{0.350pt}{0.400pt}}
\put(572,234.67){\rule{3.373pt}{0.400pt}}
\multiput(572.00,235.17)(7.000,-1.000){2}{\rule{1.686pt}{0.400pt}}
\put(586,233.17){\rule{3.100pt}{0.400pt}}
\multiput(586.00,234.17)(8.566,-2.000){2}{\rule{1.550pt}{0.400pt}}
\multiput(601.58,228.32)(0.494,-1.305){25}{\rule{0.119pt}{1.129pt}}
\multiput(600.17,230.66)(14.000,-33.658){2}{\rule{0.400pt}{0.564pt}}
\put(615,195.67){\rule{3.614pt}{0.400pt}}
\multiput(615.00,196.17)(7.500,-1.000){2}{\rule{1.807pt}{0.400pt}}
\put(630,194.17){\rule{2.900pt}{0.400pt}}
\multiput(630.00,195.17)(7.981,-2.000){2}{\rule{1.450pt}{0.400pt}}
\put(456.0,286.0){\rule[-0.200pt]{3.613pt}{0.400pt}}
\sbox{\plotpoint}{\rule[-0.500pt]{1.000pt}{1.000pt}}%
\sbox{\plotpoint}{\rule[-0.200pt]{0.400pt}{0.400pt}}%
\put(484,424){\makebox(0,0)[r]{quasi}}
\sbox{\plotpoint}{\rule[-0.500pt]{1.000pt}{1.000pt}}%
\multiput(504,424)(20.756,0.000){5}{\usebox{\plotpoint}}
\put(604,424){\usebox{\plotpoint}}
\put(211,357){\usebox{\plotpoint}}
\put(211.00,357.00){\usebox{\plotpoint}}
\put(230.00,349.00){\usebox{\plotpoint}}
\put(249.45,342.28){\usebox{\plotpoint}}
\put(269.14,335.98){\usebox{\plotpoint}}
\put(289.41,331.86){\usebox{\plotpoint}}
\put(309.02,325.07){\usebox{\plotpoint}}
\put(323.58,310.76){\usebox{\plotpoint}}
\put(340.65,299.21){\usebox{\plotpoint}}
\put(358.29,289.62){\usebox{\plotpoint}}
\put(371.19,273.66){\usebox{\plotpoint}}
\multiput(384,270)(4.956,-20.155){3}{\usebox{\plotpoint}}
\put(405.61,207.11){\usebox{\plotpoint}}
\put(425.96,203.27){\usebox{\plotpoint}}
\put(440.76,189.33){\usebox{\plotpoint}}
\put(460.44,184.93){\usebox{\plotpoint}}
\put(480.08,179.35){\usebox{\plotpoint}}
\put(500.29,174.92){\usebox{\plotpoint}}
\put(520.48,170.57){\usebox{\plotpoint}}
\put(541.21,170.00){\usebox{\plotpoint}}
\put(559.33,161.00){\usebox{\plotpoint}}
\put(580.00,159.86){\usebox{\plotpoint}}
\put(597.48,149.82){\usebox{\plotpoint}}
\put(615.12,139.00){\usebox{\plotpoint}}
\put(635.81,138.17){\usebox{\plotpoint}}
\put(644,137){\usebox{\plotpoint}}
\sbox{\plotpoint}{\rule[-0.200pt]{0.400pt}{0.400pt}}%
\put(211.0,131.0){\rule[-0.200pt]{0.400pt}{90.337pt}}
\put(211.0,131.0){\rule[-0.200pt]{104.310pt}{0.400pt}}
\put(644.0,131.0){\rule[-0.200pt]{0.400pt}{90.337pt}}
\put(211.0,506.0){\rule[-0.200pt]{104.310pt}{0.400pt}}
\end{picture}
\setlength{\unitlength}{0.240900pt}
\ifx\plotpoint\undefined\newsavebox{\plotpoint}\fi
\sbox{\plotpoint}{\rule[-0.200pt]{0.400pt}{0.400pt}}%
\begin{picture}(705,630)(0,0)
\sbox{\plotpoint}{\rule[-0.200pt]{0.400pt}{0.400pt}}%
\put(211.0,131.0){\rule[-0.200pt]{4.818pt}{0.400pt}}
\put(191,131){\makebox(0,0)[r]{$0.0001$}}
\put(624.0,131.0){\rule[-0.200pt]{4.818pt}{0.400pt}}
\put(211.0,187.0){\rule[-0.200pt]{2.409pt}{0.400pt}}
\put(634.0,187.0){\rule[-0.200pt]{2.409pt}{0.400pt}}
\put(211.0,220.0){\rule[-0.200pt]{2.409pt}{0.400pt}}
\put(634.0,220.0){\rule[-0.200pt]{2.409pt}{0.400pt}}
\put(211.0,244.0){\rule[-0.200pt]{2.409pt}{0.400pt}}
\put(634.0,244.0){\rule[-0.200pt]{2.409pt}{0.400pt}}
\put(211.0,262.0){\rule[-0.200pt]{2.409pt}{0.400pt}}
\put(634.0,262.0){\rule[-0.200pt]{2.409pt}{0.400pt}}
\put(211.0,277.0){\rule[-0.200pt]{2.409pt}{0.400pt}}
\put(634.0,277.0){\rule[-0.200pt]{2.409pt}{0.400pt}}
\put(211.0,289.0){\rule[-0.200pt]{2.409pt}{0.400pt}}
\put(634.0,289.0){\rule[-0.200pt]{2.409pt}{0.400pt}}
\put(211.0,300.0){\rule[-0.200pt]{2.409pt}{0.400pt}}
\put(634.0,300.0){\rule[-0.200pt]{2.409pt}{0.400pt}}
\put(211.0,310.0){\rule[-0.200pt]{2.409pt}{0.400pt}}
\put(634.0,310.0){\rule[-0.200pt]{2.409pt}{0.400pt}}
\put(211.0,319.0){\rule[-0.200pt]{4.818pt}{0.400pt}}
\put(191,319){\makebox(0,0)[r]{$0.001$}}
\put(624.0,319.0){\rule[-0.200pt]{4.818pt}{0.400pt}}
\put(211.0,375.0){\rule[-0.200pt]{2.409pt}{0.400pt}}
\put(634.0,375.0){\rule[-0.200pt]{2.409pt}{0.400pt}}
\put(211.0,408.0){\rule[-0.200pt]{2.409pt}{0.400pt}}
\put(634.0,408.0){\rule[-0.200pt]{2.409pt}{0.400pt}}
\put(211.0,431.0){\rule[-0.200pt]{2.409pt}{0.400pt}}
\put(634.0,431.0){\rule[-0.200pt]{2.409pt}{0.400pt}}
\put(211.0,450.0){\rule[-0.200pt]{2.409pt}{0.400pt}}
\put(634.0,450.0){\rule[-0.200pt]{2.409pt}{0.400pt}}
\put(211.0,464.0){\rule[-0.200pt]{2.409pt}{0.400pt}}
\put(634.0,464.0){\rule[-0.200pt]{2.409pt}{0.400pt}}
\put(211.0,477.0){\rule[-0.200pt]{2.409pt}{0.400pt}}
\put(634.0,477.0){\rule[-0.200pt]{2.409pt}{0.400pt}}
\put(211.0,488.0){\rule[-0.200pt]{2.409pt}{0.400pt}}
\put(634.0,488.0){\rule[-0.200pt]{2.409pt}{0.400pt}}
\put(211.0,497.0){\rule[-0.200pt]{2.409pt}{0.400pt}}
\put(634.0,497.0){\rule[-0.200pt]{2.409pt}{0.400pt}}
\put(211.0,506.0){\rule[-0.200pt]{4.818pt}{0.400pt}}
\put(191,506){\makebox(0,0)[r]{$0.01$}}
\put(624.0,506.0){\rule[-0.200pt]{4.818pt}{0.400pt}}
\put(211.0,131.0){\rule[-0.200pt]{0.400pt}{4.818pt}}
\put(211,90){\makebox(0,0){$100$}}
\put(211.0,486.0){\rule[-0.200pt]{0.400pt}{4.818pt}}
\put(298.0,131.0){\rule[-0.200pt]{0.400pt}{4.818pt}}
\put(298,90){\makebox(0,0){$120$}}
\put(298.0,486.0){\rule[-0.200pt]{0.400pt}{4.818pt}}
\put(384.0,131.0){\rule[-0.200pt]{0.400pt}{4.818pt}}
\put(384,90){\makebox(0,0){$140$}}
\put(384.0,486.0){\rule[-0.200pt]{0.400pt}{4.818pt}}
\put(471.0,131.0){\rule[-0.200pt]{0.400pt}{4.818pt}}
\put(471,90){\makebox(0,0){$160$}}
\put(471.0,486.0){\rule[-0.200pt]{0.400pt}{4.818pt}}
\put(557.0,131.0){\rule[-0.200pt]{0.400pt}{4.818pt}}
\put(557,90){\makebox(0,0){$180$}}
\put(557.0,486.0){\rule[-0.200pt]{0.400pt}{4.818pt}}
\put(644.0,131.0){\rule[-0.200pt]{0.400pt}{4.818pt}}
\put(644,90){\makebox(0,0){$200$}}
\put(644.0,486.0){\rule[-0.200pt]{0.400pt}{4.818pt}}
\put(211.0,131.0){\rule[-0.200pt]{0.400pt}{90.337pt}}
\put(211.0,131.0){\rule[-0.200pt]{104.310pt}{0.400pt}}
\put(644.0,131.0){\rule[-0.200pt]{0.400pt}{90.337pt}}
\put(211.0,506.0){\rule[-0.200pt]{104.310pt}{0.400pt}}
\put(30,318){\rotatebox{-270}{\makebox(0,0){$\frac{E-E_x}{|E_x|}$}}
}\put(427,29){\makebox(0,0){basis size $ n $}}
\put(427,568){\makebox(0,0){$ E[td\mu] $}}
\put(484,465){\makebox(0,0)[r]{pseudo}}
\put(504.0,465.0){\rule[-0.200pt]{24.090pt}{0.400pt}}
\put(211,403){\usebox{\plotpoint}}
\multiput(211.00,401.93)(1.637,-0.485){11}{\rule{1.357pt}{0.117pt}}
\multiput(211.00,402.17)(19.183,-7.000){2}{\rule{0.679pt}{0.400pt}}
\multiput(233.00,394.92)(0.814,-0.493){23}{\rule{0.746pt}{0.119pt}}
\multiput(233.00,395.17)(19.451,-13.000){2}{\rule{0.373pt}{0.400pt}}
\put(254,381.17){\rule{4.500pt}{0.400pt}}
\multiput(254.00,382.17)(12.660,-2.000){2}{\rule{2.250pt}{0.400pt}}
\multiput(276.00,379.93)(1.418,-0.488){13}{\rule{1.200pt}{0.117pt}}
\multiput(276.00,380.17)(19.509,-8.000){2}{\rule{0.600pt}{0.400pt}}
\multiput(298.58,368.79)(0.496,-1.152){39}{\rule{0.119pt}{1.014pt}}
\multiput(297.17,370.89)(21.000,-45.895){2}{\rule{0.400pt}{0.507pt}}
\multiput(319.00,323.92)(1.015,-0.492){19}{\rule{0.900pt}{0.118pt}}
\multiput(319.00,324.17)(20.132,-11.000){2}{\rule{0.450pt}{0.400pt}}
\multiput(341.00,312.93)(1.637,-0.485){11}{\rule{1.357pt}{0.117pt}}
\multiput(341.00,313.17)(19.183,-7.000){2}{\rule{0.679pt}{0.400pt}}
\multiput(363.00,305.93)(2.269,-0.477){7}{\rule{1.780pt}{0.115pt}}
\multiput(363.00,306.17)(17.306,-5.000){2}{\rule{0.890pt}{0.400pt}}
\multiput(384.00,300.93)(1.637,-0.485){11}{\rule{1.357pt}{0.117pt}}
\multiput(384.00,301.17)(19.183,-7.000){2}{\rule{0.679pt}{0.400pt}}
\multiput(406.00,293.93)(1.937,-0.482){9}{\rule{1.567pt}{0.116pt}}
\multiput(406.00,294.17)(18.748,-6.000){2}{\rule{0.783pt}{0.400pt}}
\multiput(428.58,286.29)(0.496,-0.692){39}{\rule{0.119pt}{0.652pt}}
\multiput(427.17,287.65)(21.000,-27.646){2}{\rule{0.400pt}{0.326pt}}
\multiput(449.00,258.92)(0.549,-0.496){37}{\rule{0.540pt}{0.119pt}}
\multiput(449.00,259.17)(20.879,-20.000){2}{\rule{0.270pt}{0.400pt}}
\put(492,238.17){\rule{4.500pt}{0.400pt}}
\multiput(492.00,239.17)(12.660,-2.000){2}{\rule{2.250pt}{0.400pt}}
\multiput(514.00,236.93)(1.937,-0.482){9}{\rule{1.567pt}{0.116pt}}
\multiput(514.00,237.17)(18.748,-6.000){2}{\rule{0.783pt}{0.400pt}}
\multiput(536.00,230.93)(1.560,-0.485){11}{\rule{1.300pt}{0.117pt}}
\multiput(536.00,231.17)(18.302,-7.000){2}{\rule{0.650pt}{0.400pt}}
\multiput(557.00,223.94)(3.113,-0.468){5}{\rule{2.300pt}{0.113pt}}
\multiput(557.00,224.17)(17.226,-4.000){2}{\rule{1.150pt}{0.400pt}}
\put(471.0,240.0){\rule[-0.200pt]{5.059pt}{0.400pt}}
\multiput(601.00,219.93)(2.269,-0.477){7}{\rule{1.780pt}{0.115pt}}
\multiput(601.00,220.17)(17.306,-5.000){2}{\rule{0.890pt}{0.400pt}}
\multiput(622.00,214.95)(4.704,-0.447){3}{\rule{3.033pt}{0.108pt}}
\multiput(622.00,215.17)(15.704,-3.000){2}{\rule{1.517pt}{0.400pt}}
\put(579.0,221.0){\rule[-0.200pt]{5.300pt}{0.400pt}}
\sbox{\plotpoint}{\rule[-0.500pt]{1.000pt}{1.000pt}}%
\sbox{\plotpoint}{\rule[-0.200pt]{0.400pt}{0.400pt}}%
\put(484,424){\makebox(0,0)[r]{quasi}}
\sbox{\plotpoint}{\rule[-0.500pt]{1.000pt}{1.000pt}}%
\multiput(504,424)(20.756,0.000){5}{\usebox{\plotpoint}}
\put(604,424){\usebox{\plotpoint}}
\put(211,350){\usebox{\plotpoint}}
\multiput(211,350)(12.274,-16.737){2}{\usebox{\plotpoint}}
\put(236.96,318.30){\usebox{\plotpoint}}
\put(256.18,310.60){\usebox{\plotpoint}}
\multiput(276,307)(20.421,-3.713){2}{\usebox{\plotpoint}}
\put(317.67,301.13){\usebox{\plotpoint}}
\put(335.05,290.06){\usebox{\plotpoint}}
\put(354.43,284.17){\usebox{\plotpoint}}
\put(373.51,277.00){\usebox{\plotpoint}}
\put(392.64,270.21){\usebox{\plotpoint}}
\multiput(406,269)(14.347,-14.999){2}{\usebox{\plotpoint}}
\put(438.95,232.97){\usebox{\plotpoint}}
\put(454.13,221.00){\usebox{\plotpoint}}
\put(474.78,220.10){\usebox{\plotpoint}}
\put(494.94,215.20){\usebox{\plotpoint}}
\multiput(514,210)(11.275,-17.426){2}{\usebox{\plotpoint}}
\put(538.02,176.00){\usebox{\plotpoint}}
\multiput(557,176)(17.869,-10.559){2}{\usebox{\plotpoint}}
\put(593.03,152.16){\usebox{\plotpoint}}
\put(611.27,143.07){\usebox{\plotpoint}}
\put(631.60,140.00){\usebox{\plotpoint}}
\put(644,140){\usebox{\plotpoint}}
\sbox{\plotpoint}{\rule[-0.200pt]{0.400pt}{0.400pt}}%
\put(211.0,131.0){\rule[-0.200pt]{0.400pt}{90.337pt}}
\put(211.0,131.0){\rule[-0.200pt]{104.310pt}{0.400pt}}
\put(644.0,131.0){\rule[-0.200pt]{0.400pt}{90.337pt}}
\put(211.0,506.0){\rule[-0.200pt]{104.310pt}{0.400pt}}
\end{picture}
\caption{The ground state energy $E$ of the Ps$^-$ (left)
and $td\mu$ (right) systems
as function of the size $n$ of the correlated
Gaussian bases chosen stochastically
with pseudo- or quasi-random sequences.
The exact energies are
$E_x=-0.26200507$ for Ps$^-$ and
$E_x=-111.36444$ for $td\mu$~\cite{suzuki-varga}.
	}
	\label{fig-basis} \end{figure}
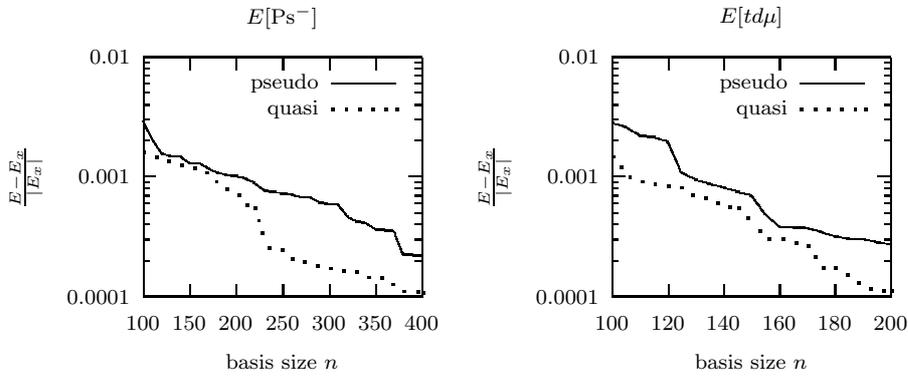

Figure~(\ref{fig-basis}) compares the ground-state energies
of these systems calculated using correlated Gaussian bases constructed
stochastically with pseudo- or quasi-random sequences.
For all basis sizes
the quasi-random sequence
provides lower variational estimate
of the ground-state energy
than
the pseudo-random sequence.

\section{Conclusion}
In the Correlated Gaussian method the variational basis if often
constructed stochastically using pseudo-random sequences.  Pseudo-random
sequences have the intrinsic property of high discrepancy: they produce
relatively large non-sampled areas.  On the contrary, the low-discrepancy
(quasi-random) sequences, like the Van der Corput sequence, are
specifically designed to have low discrepancy.

We have compared the relative performance of the two types of sequences
by calculating the ground-state energies of two Coulombic three-body
systems using stochastic correlated Gaussian bases of various sizes built
with both pseudo- and quasi-random sequences.  For all bases sizes the
quasi-random sequence outperforms the pseudo-random sequence by providing
a lower variational estimate of the energy.

\end{document}